\documentclass[conference]{IEEEtran}
\IEEEoverridecommandlockouts
\usepackage{cite}
\usepackage{amsmath,amssymb,amsfonts}
\usepackage{algorithmic}
\usepackage{graphicx}
\usepackage{textcomp}
\usepackage{xcolor}
\usepackage{glossaries}
\usepackage[hyphens]{url}
\usepackage{hyperref}
\usepackage[noabbrev,capitalise]{cleveref}
\usepackage{orcidlink}
\usepackage[separate-uncertainty=true]{siunitx}
\usepackage{booktabs}
\usepackage{multirow}
\usepackage{makecell}
\usepackage{threeparttable}

\glsenableentrycount

\newacronym[longplural=systems-on-chip]{soc}{SoC}{system-on-chip}
\newacronym{isa}{ISA}{instruction set architecture}
\newacronym{pulp}{PULP}{parallel ultra-low power}
\newacronym{odrg}{ODRG}{on-demand redundancy grouping}
\newacronym{tcls}{TCLS}{triple-core lockstep}
\newacronym{ecc}{ECC}{error correction codes}
\newacronym{dut}{DUT}{device under test}
\newacronym{wdt}{WDT}{watchdog timer}
\newacronym{seu}{SEU}{single-event upset}
\newacronym{see}{SEE}{single-event effect}
\newacronym{sefi}{SEFI}{single-event functional interrupt}
\newacronym{set}{SET}{single-event transient}
\newacronym{sel}{SEL}{single-event latchup}
\newacronym{let}{LET}{linear energy transfer}
\newacronym{sdc}{SDC}{silent data corruption}
\newacronym{due}{DUE}{detectable unrecoverable error}
\newacronym{tid}{TID}{total ionizing dose}
\newacronym{rhbd}{RHBD}{radiation hardening by design}
\newacronym{ge}{GE}{gate equivalent}
\newacronym{udma}{$\mu$DMA}{I/O DMA}
\newacronym{gpio}{GPIO}{general purpose input/output}
\newacronym{dma}{DMA}{direct memory access}
\newacronym{tcdm}{TCDM}{tightly coupled data memory}
\newacronym{pdk}{PDK}{process desing kit}
\newacronym{pcb}{PCB}{printed circuit board}
\newacronym{sram}{SRAM}{static random-access memory}
\newacronym{tmr}{TMR}{triple modular redundancy}
\newacronym{dmr}{DMR}{dual modular redundancy}
\newacronym{secded}{SECDED}{single error correction, double error detection}
\newacronym{mftf}{MFTF}{mean fluence to failure}
\newacronym{mttf}{MTTF}{mean time to failure}
\newacronym{rpi}{RPi}{Raspberry Pi}
\newacronym{ff}{FF}{flip-flop}
\newacronym{fpga}{FPGA}{field-programmable gate array}
\newacronym{fec}{FEC}{forwards error correction}
\newacronym{obi}{OBI}{Open Bus Interface}
\newacronym{axi}{AXI5}{Advanced eXtensible Interface 5}
\newacronym{apb}{APB5}{Advanced Peripheral Bus 5}
\newacronym[longplural=networks-on-chip]{noc}{NoC}{network-on-chip}
\newacronym{fo4}{FO4}{fan-out of 4 delay}
\newacronym{tmrg}{TMRG}{Triple Modular Redundancy Generator}

\def\BibTeX{{\rm B\kern-.05em{\sc i\kern-.025em b}\kern-.08em
    T\kern-.1667em\lower.7ex\hbox{E}\kern-.125emX}}
\begin{document}
\bstctlcite{IEEEexample:BSTcontrol}

\title{relOBI: A Reliable Low-latency Interconnect for Tightly-Coupled On-chip Communication
\thanks{This work has received funding from the Swiss State Secretariat for Education, Research, and Innovation (SERI) under the SwissChips initiative.}
}

\author{\IEEEauthorblockN{Michael Rogenmoser}
\IEEEauthorblockA{\textit{IIS, ETH Zurich}\\
Zurich, Switzerland\\
michaero@iis.ee.ethz.ch\,\orcidlink{0000-0003-4622-4862}}
\and
\IEEEauthorblockN{Angelo Garofalo}
\IEEEauthorblockA{\textit{IIS, ETH Zurich}\\
Zurich, Switzerland\\
\textit{DEIS, University of Bologna}\\
Bologna, Italy\\
angelo.garofalo@unibo.it\,\orcidlink{https://orcid.org/0000-0002-7495-6895}
}
\and
\IEEEauthorblockN{Luca Benini}
\IEEEauthorblockA{\textit{IIS, ETH Zurich}\\
Zurich, Switzerland\\
\textit{DEIS, University of Bologna}\\
Bologna, Italy\\
lbenini@iis.ee.ethz.ch\,\orcidlink{0000-0001-8068-3806}
}
}

\maketitle

\begin{abstract}
On-chip communication is a critical element of modern \cglspl{soc}, allowing processor cores to interact with memory and peripherals. Interconnects require special care in radiation-heavy environments, as any soft error within the \cgls{soc} interconnect is likely to cause a functional failure of the whole \cgls{soc}. This work proposes \textit{relOBI}, an extension to the \textit{\gls{obi}} combining \cgls{tmr} for critical handshake signals with \cgls{ecc} protection on other signals. Implementing and testing the reliable crossbar shows improved reliability to injected single faults from a vulnerability of \SI{34.85}{\percent} to zero compared to the irredundant baseline, with an area increase of \SI{2.6}{\times}. The area overhead is \SI{1.8}{\times} lower than that reported in the literature for fine-grained triplication and voting.
\end{abstract}

\glsresetall
% \glsreset{ecc}


\section{Introduction}

% TODO problem statement
As more satellites are deployed, especially in constellations for earth observation or communication~\cite{kulu_satellite_2024}, many more advanced \cglspl{soc} are being used in the hostile space environment. Radiation-induced \cglspl{seu} and \cglspl{set} pose significant threats to system reliability, particularly in crucial devices, components, and applications where high availability and fault tolerance are paramount. Smaller, more modern technology nodes are more likely to be affected by \cglspl{set}, as increasing clock frequency and shrinking transistors make it more likely that transient faults are latched by sequential elements instead of being masked thanks to temporal masking~\cite{di_mascio_open-source_2021}. We target soft errors such as \cglspl{seu} and \cglspl{set}, assuming any net driven by a transistor can flip. \cGlspl{seu} are modeled to directly affect the state of a register until overwritten, while \cglspl{set} are applied long enough such that they are clocked into all affected registers for evaluation.

Recent testing of a partially protected microcontroller design under radiation and in simulation revealed that faults in high activity areas, such as the system interconnect, severely impact overall device reliability~\cite{rogenmoser_trikarenos_2025}. While processor cores and memory banks are %already protected in~\cite{rogenmoser_trikarenos_2025}, the interconnect presents a single point of failure, where its vulnerability highlights the importance of ensuring this part is tolerant to soft errors.
usually protected~\cite{rogenmoser_trikarenos_2025}, protecting the interconnect is also critical, but interconnect protection approaches are less widely explored and understood.

Previous research has explored methods for protecting large \cglspl{noc}~\cite{bertozzi_error_2005, radetzki_methods_2013, bhamidipati_hren_2022}, where the main options proposed are error correction encoding, error checking with retransmission, and ensuring reliability at the transistor level.
However, these methods and their implementation mostly rely on offline test modes, which disrupt normal operation and cannot fully counter soft errors. Furthermore, these \cgls{noc}-centric approaches have not been adapted to smaller, tightly coupled low-latency interconnects.
% The authors in~\cite{andorno_radiation-tolerant_2023} enhance the reliability of a microcontroller system with 
Focusing on low-latency interconnects for microcontrollers, \cite{andorno_radiation-tolerant_2023}~enhances reliability with fine-grained triplication and voting for a \SIrange[range-units=single, range-phrase=-]{4.8}{6.9}{\times} area increase and \SI{1.4}{\times} critical path timing increase. A lighter protection is applied to the \cgls{apb}~\cite{arm_amba-apb_2021} by triplicating only the control and protecting data and address with \cgls{ecc}, reducing the number of signals by \SI{42}{\percent} w.r.t. full triplication~\cite{andorno_radiation-tolerant_2023}.
% Full triplication of a processor core and the interconnection network is explored in~\cite{andorno_radiation-tolerant_2023} for a \SIrange{4.8}{6.9}{\times} increase in area utilization, with a lighter protection approach on a simple \cgls{apb} with \cgls{ecc}-protected data and address, and triplicated control, reducing signals by \SI{42}{\percent}.
% However, tightly-coupled interconnection networks, commonly used as local interconnects within processor and accelerator tiles, have not been explored in depth. 
The \textit{\cgls{obi}}~\cite{bink_obi_2023} and \textit{\cgls{axi}}~\cite{arm_amba_2023} standards define integrity signalling as an optional extension, which adds parity to various signals for error detection, but the specifications do not detail a correction or recovery mechanism. Retransmission may correct some error types, but handshake errors are not recoverable, meaning a system reset would be required. Relying on the underlying transistors and cells to be tolerant can address this issue, but is limited to technologies where cell implementation has been radiation hardened.

% however, tightly-coupled interconnection networks, commonly used as local interconnects within processor and accelerator tiles, have not been explored as deeply. The main options proposed for \cgls{noc} are corrective encoding, error checking with retransmission, and ensuring reliability on the transistor level. The \textit{\cgls{obi}}~\cite{bink_obi_2023} and \textit{\cgls{axi}}~\cite{arm_amba_2023} standards define integrity signalling as an optional extension, which adds parity to various signals, but does not specify a recovery mechanism. 
% Retransmission may be an option for some error types, but handshake errors are not recoverable, meaning a system reset would be required. Relying on the underlying transistors and cells to be tolerant can enable good results, but limits applicability to technologies where this is available and may not account for manufacturing defects.

% Luca's NoC reliability paper~\cite{bertozzi_low_2002, bertozzi_error_2005}

% Relevant survey~\cite{radetzki_methods_2013}

% TODO contributions

To address these limitations, we present \textit{relOBI}, a novel augmentation of the on-chip interface \textit{\cgls{obi}}~\cite{bink_obi_2023} that provides complete soft error tolerance for combinatorial and sequential elements within system interconnects and allows for continuous operation at a much lower cost than fine-grained triplication. Specifically, this work includes the following contributions:

\begin{itemize}
    \item Definition of an extension to the \textit{\cgls{obi}} interface allowing for rapid encoding and decoding to handle any single error on the interface wires while maintaining flexibility.
    \item Design and implementation of open-source interconnect IP blocks\footnote{\label{githuburl}\url{https://github.com/pulp-platform/obi/tree/relOBI}}, including a crossbar architecture, that ensure tolerance to \cglspl{seu} and \cglspl{set} for correct transaction handling and routing.
    \item Evaluation of the performance and area implications of the proposed interconnect, comparing to an \textit{\cgls{obi}} baseline and the \textit{\cgls{obi}} baseline augmented with fine-grained triplication for reliability.
    \item Soft error tolerance analysis and assessment of the proposed interface extension and interconnect IP blocks through fault simulation, demonstrating the approach's effectiveness.
\end{itemize}

% The novel \textit{relOBI} design shows significantly smaller area impact than fine-grained triplication at a similar timing penalty, while still ensuring full tolerance to \cglspl{seu} allowing for continuous operation.
% add a clear statement of novelty wrt the soa: much reduced overhead wrt to fine grained schemes and full robustness compared to the noc schemes that do no  protect everything.

\section{Architecture}

\begin{figure}[t]
    \centering
    \includegraphics[width=\columnwidth]{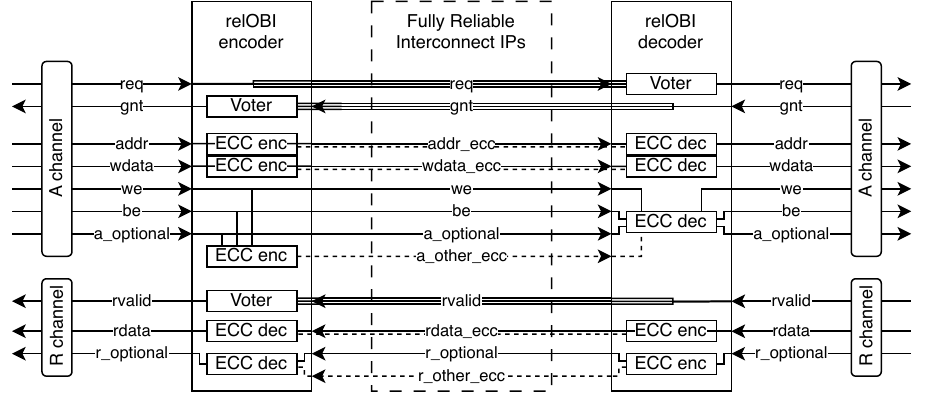}
    \caption{Encoding and decoding between \textit{\cgls{obi}} and \textit{relOBI}, showing triplicated handshake signals and \cgls{ecc} protection for other signals.}
    \label{fig:encoding}
\end{figure}

To ensure correct handling of \textit{\cgls{obi}} transactions, all handshaking signals are triplicated and voted, as shown in \Cref{fig:encoding}. 
This ensures that both the manager and subordinate keep a consistent view of the interconnect to avoid the creation or deletion of a transaction due to a soft error.
%This ensures that any single error on these signals does not affect the state of the manager and subordinate of the interfaces differently. 
Furthermore, all other signals, such as address (\texttt{addr}), read and write data (\texttt{rdata}, \texttt{wdata}), write enable (\texttt{we}), and byte enable (\texttt{be}), are protected with \cgls{ecc}. To balance overheads in signalling and en-/decoding, data and address signals are protected individually while the remaining signals are grouped per direction, as shown in \Cref{fig:encoding}. This protection scheme ensures continuous operation at no additional latency and enables in-flight correction. Moreover, error protection schemes, such as parity and retransmission, cannot easily support atomic transactions, which require a single consistent state to be kept; in contrast, the implemented forward error correction seamlessly supports such features. The encoding and decoding blocks enable compatibility with existing \textit{\cgls{obi}} interfaces.% and are assumed to be protected with other methods.

Additional modules are required to interconnect different managers and subordinates properly. To access multiple subordinates from a single manager, this manager is connected to the subordinate port of an \textit{\cgls{obi}} demultiplexer module, which properly routes the requests and ensures the strict ordering requirements in \textit{\cgls{obi}}~\cite{bink_obi_2023} are enforced. An \textit{\cgls{obi}} multiplexer module allows multiple managers to access a single subordinate, providing a fair round-robin arbitration between the requesting managers and providing backpressure when there is contention. An internal FIFO keeps track of the requests sent to match them to the responses and properly select the response port. Combining these modules to create a crossbar architecture allows multiple managers to connect to multiple subordinates, allowing different requests to bypass each other if the managers and subordinates are independent.

\begin{figure}[t]
    \centering
    \includegraphics[width=\linewidth]{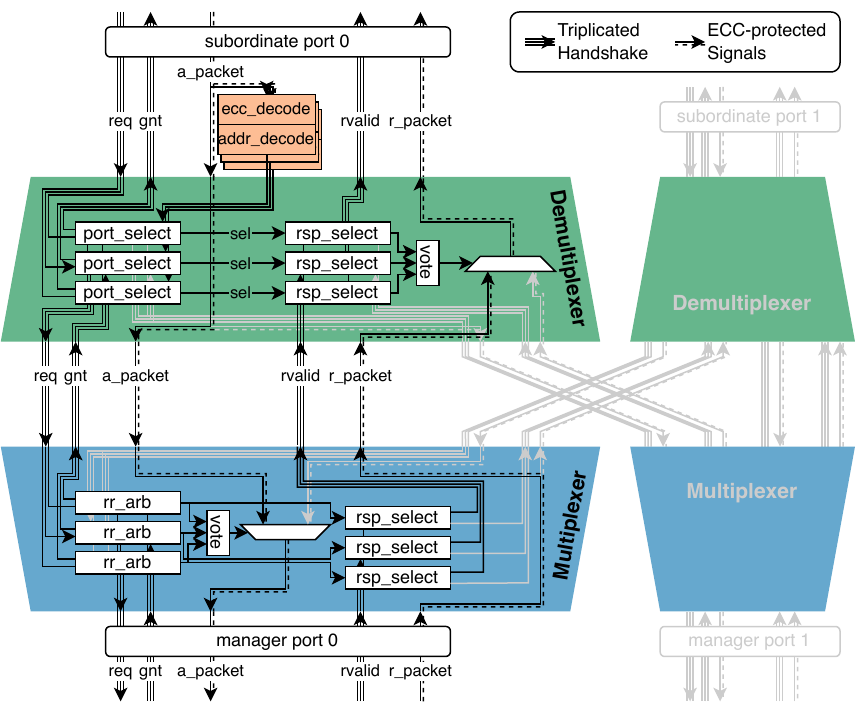}
    \caption{Microarchitecture of a reliable \textit{relOBI} crossbar, comprising demultiplexer (top, green) and multiplexer (bottom, blue) blocks and address decoders to select the target port (orange). For this 2$\times$2 crossbar, the secondary ports and modules are grey for clarity, however they have the same structure as the primary ports and modules.}
    \label{fig:mux_crossbar}
\end{figure}

These interconnect blocks are internally hardened to tolerate \cglspl{seu} and \cglspl{set}: The modules are protected by ensuring that all handshake-controlling signals and the corresponding logic is triplicated, as shown for the crossbar in \Cref{fig:mux_crossbar}. Each of the three handshakes is controlled individually with the corresponding logic from the reference \textit{\cgls{obi}} blocks. The internal state of these control blocks is synchronized across the blocks with voters directly following the registers, with this voted signal used for the following circuitry, correcting latent errors in these registers. Where applicable, \cgls{ecc}-protected packets are passed through to avoid triplication overhead, as the individual bits are already protected. However, if any information inside the \cgls{ecc}-protected signals is required, such as the address to determine the subordinate port (see orange blocks in \Cref{fig:mux_crossbar}), this signal is triplicated and properly decoded individually to ensure in-flight correction, even if the decoder is affected by a \cgls{set}. As the control logic is already triplicated, these individual signals are independently handled by the corresponding logic. If selection of an \cgls{ecc}-protected signal is required inside a \textit{relOBI} module, e.g., for the write data in the \textit{relOBI} multiplexer or the read data in the \textit{relOBI} demultiplexer, a simple multiplexer is used. However, as the control signal is triplicated, each bit in the multiplexer has an individually voted control signal. This ensures that a \cgls{set} in a voter can be corrected by the \cgls{ecc} protection, as only a single bit will end up with an incorrect selection.

% However, all decoding, modification, and selection of these signals is triplicated and voted to ensure correctness.
% \textcolor{red}{TODO explain fully reliable interconnect IPs, impact on blocks, ...}

% \begin{figure}[t]
%     \centering
%     \includegraphics[width=\linewidth]{fig/relobi-crossbar2.drawio.pdf}
%     \caption{Microarchitecture of a reliable \textit{relOBI} crossbar, comprising demultiplexer (top) and multiplexer (bottom) blocks and address decoders to select the target port. For this 2$\times$2 crossbar, the secondary ports and modules are grey for clarity, however they have the same structure as the primary ports and modules.}
%     \label{fig:mux_crossbar}
% \end{figure}

\begin{table*}[t]
    \centering
    \caption{Fault injection simulation results for the baseline \textit{\cgls{obi}}, reliable \textit{relOBI}, and \acrshort{tmrg} \textit{\cgls{obi}} crossbar implementations. Each configuration is simulated under different fault conditions for \SI{1000000}{} faults, a subset of the total possible faults.}
    \begin{threeparttable}
    \begin{tabular}{@{}ll@{ }lS[table-format=2.2, table-space-text-post={\,\si{\percent}}]<{\,\si{\percent}}S[table-format=2.2, table-space-text-post={\,\si{\percent}}]<{\,\si{\percent}}S[table-format=2.2, table-space-text-post={\,\si{\percent}}]<{\,\si{\percent}}S[table-format=2.2, table-space-text-post={\,\si{\percent}}]<{\,\si{\percent}}S[table-format=2.2, table-space-text-post={\,\si{\percent}}]<{\,\si{\percent}}rS[table-format=10.0]@{}}\toprule
        \multirow{2}{*}{\textbf{Design}} & \multicolumn{2}{c}{\multirow{2}{*}{\textbf{Fault Type}}} & \multicolumn{3}{c}{\textbf{Correct Interface Behavior}} & \multicolumn{2}{c}{\textbf{Incorrect Behavior}} & \multicolumn{1}{c}{\textbf{Total}} & \textbf{Total} \\
                        &              &      & \multicolumn{1}{c}{Masked}               & \multicolumn{1}{c}{Corrected}            & \multicolumn{1}{c}{Uncorrectable\tnote{1}}    & \multicolumn{1}{c}{Uncorrectable\tnote{1}}    & \multicolumn{1}{c}{Undetected} & \multicolumn{1}{c}{\textbf{Injected Faults}}  & \textbf{Possible Faults}\\\midrule
        \multirow{3}{*}{\makecell[l]{\textit{\cgls{obi}}\\crossbar}}    & RTL & \texttt{FLOP}           & 65.15 & \multicolumn{1}{c}{-} & \multicolumn{1}{c}{-} & \multicolumn{1}{c}{-} & 34.85 & \SI{1000000}{} (\SI{2.18}{\percent}) & 45877566 \\
            & RTL & \texttt{PORT}           & 73.06 & \multicolumn{1}{c}{-} & \multicolumn{1}{c}{-} & \multicolumn{1}{c}{-} & 26.94 & \SI{1000000}{} (\SI{0.56}{\percent}) & 178866820 \\
                & Netlist & \texttt{PRIM}       & 88.48 & \multicolumn{1}{c}{-} & \multicolumn{1}{c}{-} & \multicolumn{1}{c}{-} & 11.52 & \SI{1000000}{} (\SI{0.06}{\percent}) & 1634033170 \\\addlinespace
        \multirow{3}{*}{\makecell[l]{\textit{relOBI}\\crossbar}} & RTL & \texttt{FLOP}           & 78.08 & 21.92 & 0 & 0 & 0 & \SI{1000000}{} (\SI{1.63}{\percent}) & 61250028 \\
             & RTL & \texttt{PORT}           & 61.19 & 38.53 & 0.28 & 0 & 0   & \SI{1000000}{} (\SI{0.10}{\percent}) & 982580500 \\
                 & Netlist & \texttt{PRIM}       & 74.41 & 24.82 & 0.77 & 0 & 0   & \SI{1000000}{} (\SI{0.03}{\percent}) & 3672057400 \\\addlinespace
        \multirow{3}{*}{\makecell[l]{\acrshort{tmrg}\\\textit{\cgls{obi}}\\crossbar}} & RTL & \texttt{FLOP} & 29.59 & 70.41 & \multicolumn{1}{c}{-} & \multicolumn{1}{c}{-} & 0 & \SI{1000000}{} (\SI{0.74}{\percent}) & 134539020 \\
         & RTL & \texttt{PORT} & 70.20 & 29.80 & \multicolumn{1}{c}{-} & \multicolumn{1}{c}{-} & 0 & \SI{1000000}{} (\SI{0.15}{\percent}) & 674689290 \\
         & Netlist & \texttt{PRIM} & 67.24 & 32.76 & \multicolumn{1}{c}{-} & \multicolumn{1}{c}{-} & 0 & \SI{1000000}{} (\SI{0.01}{\percent}) & 6725958360 \\
        \bottomrule
    \end{tabular}
    \begin{tablenotes}
    \item [1] \textit{relOBI} can detect an uncorrectable \cgls{ecc} error, referenced here. This is indicated as an output signal from the module. It does not mean that an incorrect behavior of the design was measured.
    \end{tablenotes}
    \end{threeparttable}
    \label{tab:fi_results}
\end{table*}

\section{Results}

% The tested configurations incorporate six interfaces for \textit{\cgls{obi}} managers to connect to, which are directly pipelined internally. These interfaces connect to a crossbar module, demuxing the manager interfaces with static internal address mappings and muxing for the target ports. This allows for connecting eight \textit{\cgls{obi}} subordinates, which are again pipelined. For the reliable configuration, the encoder and decoder are not part of the tested blocks, as these are assumed to be separately protected by extending and reusing the protection methods applied to the manager and subordinate components, e.g., by triplicating and voting the encoding.

% Shown in~\Cref{fig:testsetup}, a
A 6$\times$8 \textit{\cgls{obi}} crossbar system is used for testing. The 6 system managers are individually connected to the subordinate ports of the crossbar, and internally to the demultiplexers (shown in green in \Cref{fig:mux_crossbar}). The latter are fully connected to the 8 multiplexers (shown in blue in \Cref{fig:mux_crossbar}), which expose the 8 manager ports towards the system subordinates. 
% A 6$\times$8 crossbar is used for testing, a module where 6 \textit{\cgls{obi}} or \textit{relOBI} managers individually connect to the subordinate ports of the demultiplexers. These are fully connected to 8 \textit{\cgls{obi}} or \textit{relOBI} multiplexers exposing 8 manager ports for 8 subordinates.
To ensure the tested crossbars are not impacted by input and output delays, each interface is pipelined inside the tested module before and after the crossbar. 

% \begin{figure}[t]
%     \centering
%     \includegraphics[width=1\columnwidth]{fig/relobi-test_setup.drawio.pdf}
%     \caption{Tested crossbar configuration}
%     \label{fig:testsetup}
% \end{figure}

As a comparison value, we implement the same crossbar with \textit{\cgls{obi}}, \textit{relOBI}, and by applying fine-grained triplication to the baseline \textit{\cgls{obi}} design using \cgls{tmrg}~\cite{kulis_single_2017}, a similar method to the one used in~\cite{andorno_radiation-tolerant_2023}. 
The \cgls{tmrg} tool was configured to triplicate the entire design and insert voters following every sequential element, ensuring any state corruption is corrected in the following cycle. This voted state was then used for the triplicated combinational logic. Code modification was required to properly process the source code, as the \cgls{tmrg} tool does not support many SystemVerilog features. 
When \textit{relOBI} is considered, the crossbar does not include the encoders and decoders within the tested module, as these components are assumed to be separately protected by extending and reusing protection methods applied to the manager and subordinate components, e.g., by triplicating and voting the encoding.

Each bus is configured with 32-bit address and data widths and optional features (e.g., atomic support), resulting in a total bus width of 137 bits. Converting to \textit{relOBI} results in a 177-bit bus width, an increase of \SI{29}{\percent} in the number of signals. This is \SI{2.3}{\times} better than triplication of all signals, as would be required for fine-grained triplication.

% Total a optional mgr: 5  +4+3+6+2+3+1=24
% Total a optional sbr: 5+3+4+3+6+2+3+1=27
% Total r optional mgr: 5  +3+1        =9
% Total r optional sbr: 5+3+3+1        =12

% Considering only manager interface
% Total signals    obi = 2+32  +32  +4+1+24  +1+32  +9  =137
% Total signals relobi = 6+32+7+32+7+4+1+24+7+3+32+7+9+6=177

\textit{Synopsys VC\_Z01X 2025.06} was used to evaluate the design's fault tolerance characteristics. A randomized testbench simulated 1000 transactions from each manager, with the subordinates simply responding. Single transient faults were injected into the crossbar internals, injecting \texttt{FLOP} (sequential blocks) and \texttt{PORT} (module ports) faults in RTL, as well as \texttt{PRIM} (Verilog primitives) faults in the synthesized netlist. The results in \Cref{tab:fi_results} illustrate whether the \textit{\cgls{obi}} interface of the crossbar implementations with injected faults behaved the same as in the reference simulation without faults. As the reliable designs support the detection of an error, additional subdivisions are added to the correct and incorrect interface behavior columns in~\Cref{tab:fi_results}, indicating any detected corrections or uncorrectable internal faults.

The baseline \textit{\cgls{obi}} crossbar shows \SIrange[range-units=single, range-phrase=-]{11.52}{34.85}{\percent} of injected faults cause errors, while the \textit{relOBI} design corrects all injected faults, with \SIrange[range-units=single, range-phrase=-]{21.92}{38.53}{\percent} indicating an internal error was corrected. The \textit{relOBI} design indicates a separate error type when the \cgls{ecc} decoding detects an uncorrectable (multi-bit) error, illustrated by the Uncorrectable columns in~\Cref{tab:fi_results}. While this signal indicates an uncorrectable fault in less than \SI{1}{percent} of injected faults in \textit{relOBI}, this never led to incorrect behavior of the design. The injected faults directly affected the indication logic in these cases, not the \cgls{ecc} protected parts of the design. The \cgls{tmrg} design also corrects all internal faults, with \SIrange[range-units=single, range-phrase=-]{29.80}{70.41}{\percent} indicating a correction was performed. 

To evaluate the physical characteristics, the design was synthesized in a commercial \SI{7}{\nano\meter} node in typical conditions with \textit{Synopsys Design Compiler NXT 2024.09} in topographical mode, ensuring triplicated partitions are kept and their boundary is not optimized to avoid removing reliability structures.
%Without timing pressure, 
Implemented at \SI{500}{\mega\hertz}, the \textit{relOBI} crossbar requires \SI{123}{\kilo GE} compared to the \SI{47}{\kilo GE} required for the \textit{\cgls{obi}} crossbar, an increase of \SI{2.6}{\times}. This is a smaller overhead than simply triplicating the entire circuit (\SI{3}{\times}), even neglecting any internal synchronization to avoid latent errors or voting at the outputs. The interconnect area in a system utilizing an \textit{\cgls{obi}} crossbar occupies less than \SI{2}{\percent} of the \cgls{soc} area~\cite{rogenmoser_trikarenos_2025}, showing that the area overhead is quite small at the system level. Applying fine-grained triplication to the \textit{\cgls{obi}} design with \cgls{tmrg} requires \SI{219}{\kilo GE}, a \SI{4.7}{\times} overhead of the reference design. Similar results are reported in~\cite{andorno_radiation-tolerant_2023}, measuring a \SIrange[range-units=single, range-phrase=-]{4.8}{6.9}{\times} overhead. This shows that the \cgls{tmrg} design requires \SI{1.8}{\times} more area than the \textit{relOBI} implementation.

\begin{figure}[t]
    \centering
    \includegraphics[width=\columnwidth]{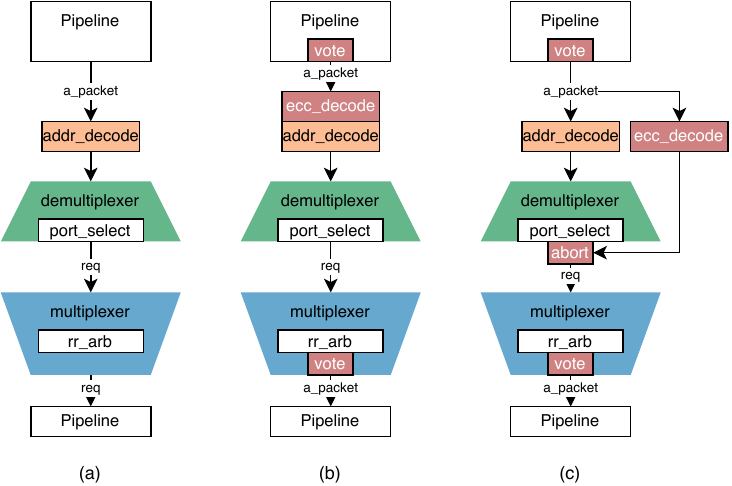}
    \caption{Diagram illustrating the critical path through the tested design, with (a) representing the critical path in the base \textit{\cgls{obi}} design, (b) in the \textit{relOBI} design, and (c) the optimized \textit{relOBI} design. Red blocks indicate the added logic for reliability.}
    \label{fig:timing}
\end{figure}

Analyzing the critical path of the design in terms of \cgls{fo4}, we see the baseline \textit{\cgls{obi}} design requires \SI{13.6}{\cgls{fo4}}. With the modifications for reliability, the \textit{relOBI} design requires \SI{23.8}{\cgls{fo4}}. Looking more closely at the critical path shown in \Cref{fig:timing}, there are two voters and the \cgls{ecc} decoder for the address added to the critical path to ensure correctness. The overhead is shown in \Cref{tab:timing}, indicating that the \SI{6.2}{\cgls{fo4}} additional delay introduced by the \cgls{ecc} decoder is the majority of the timing overhead. Modifying the circuit to check for \cgls{ecc} errors on a separate path, aborting the request if an error is detected and requiring the pipeline register to incorporate correction logic, reduces the critical path to \SI{19.1}{\cgls{fo4}}, with the delays traceable to the added voting logic and the additional gate to abort the request, as shown in \Cref{fig:timing} and \Cref{tab:timing}. The \cgls{tmrg} implementation achieves an optimal timing of \SI{16.7}{\cgls{fo4}}, a \SI{1.23}{\times} overhead. The single voter added following the sequential elements, as well as the increase in the required design area, accounts for the timing overhead. As only a single voter is required, the \cgls{tmrg} configuration shows a \SI{0.87}{\times} shorter critical path than \textit{relOBI}.

% With the internal reliability, the critical timing through the interconnect increases from \SI{13.6}{\cgls{fo4}} to \SI{23.8}{\cgls{fo4}}, due to added voters and \cgls{ecc} decoding to ensure correct routing. Modifying the design to abort transactions on an \cgls{ecc} error and recovering in the next cycle reduces the achievable clock period to \SI{19.1}{\cgls{fo4}}, similar to a fine-grained triplication~\cite{andorno_radiation-tolerant_2023}.

While the module-level relative timing overhead of \textit{relOBI} is larger than \cgls{tmrg}, the critical path is quite short in the absolute sense, and would bound the frequency of the design above \SI{2}{\giga\hertz}. Hence, when \textit{relOBI} is integrated in a design, the critical path would most likely be determined by the manager and subordinate components, such as processor cores and memory banks connected through the crossbar.

% TODO do synthesis and explain synthesis results, due to many internal hierarchies regrouping is critical, 
% GF12: going from \SI{48}{\kilo GE} to \SI{105}{\kilo GE}, an increase of \SI{119}{\percent}. 
% TSMC7: 
% going from \SI{36}{\kilo GE} to \SI{84}{\kilo GE}, an increase of \SI{132}{\percent}.

% \textcolor{red}{comment on results}

% \begin{table}[t]
%     \centering
%     \caption{Overall results comparison}
%     \begin{tabular}{@{}lrrr@{}}\toprule
%         Design       & Area [\si{\kilo GE}] & Timing [\si{\cgls{fo4}}] & Vulnerability \\\midrule
%         OBI baseline & \\
%         relOBI       & \\
%         OBI TMR      & \\\bottomrule
%     \end{tabular}
%     \label{tab:placeholder}
% \end{table}

\section{Conclusion}

The proposed \textit{relOBI} interface and protected interconnection blocks are shown to be fully \cgls{seu}-tolerant, reducing the vulnerability from up to \SI{34.85}{\percent} to zero at a much lower area impact compared to fine-grained triplication.
In the context of a reliable microcontroller design~\cite{rogenmoser_trikarenos_2025}, where cores and memories are protected, \textit{relOBI} would significantly increase system-level \cgls{seu} coverage, at a negligible area and timing overhead.  The \textit{relOBI} design is available fully open-source\footref{githuburl}.
% If applied to a microcontroller interconnect such as in Trikarenos~\cite{rogenmoser_trikarenos_2025}, both area and timing overheads would remain minimal while the complete \cgls{soc} fault tolerance would be significantly improved.
% While the standalone synthesis of the design shows an increase in area and timing, no additional latency is introduced by the design.

\begin{table}[t]
    \centering
    \caption{Timing overhead of the critical path in \cgls{fo4} for different stages of the path, comparing the \textit{relOBI} and optimized \textit{relOBI} design to the baseline \textit{\cgls{obi}} design. $\Delta$ indicates the overhead for that stage compared to the \textit{\cgls{obi}} reference: $\Delta = (T_{i}-T_{i-1}) - (T_{ref,i}-T_{ref,i-1})$.}
    \begin{tabular}{@{}l@{}rr@{}rr@{}rr@{}r@{}}\toprule
        \multirow{2}{*}{Timing stage}      & \multicolumn{1}{c}{\textit{\cgls{obi}}} & \multicolumn{2}{c}{\multirow{2}{*}{\textit{relOBI} ($\Delta$)}} &  \multicolumn{2}{c}{optimized} & \multicolumn{2}{c}{\cgls{tmrg}} \\
         & \multicolumn{1}{c}{reference} & & & \multicolumn{2}{c}{\textit{relOBI} ($\Delta$)} & \multicolumn{2}{c}{\textit{\cgls{obi}} ($\Delta$)}\\\midrule
        pipeline out      & 2.9\hspace{0.3cm}  & 4.2 & (+1.3) & 4.9 & (+2.0) & 4.4 & (+1.5)\\
        addr decode out   & 5.6\hspace{0.3cm}  & 13.1 & (+6.2) & 7.4 & (-0.2) & 7.2 & (+0.1) \\
        demultiplexer out & 7.9\hspace{0.3cm}  & 15.6 & (+0.2) & 11.1 & (+1.4) & 9.8 & (+0.3) \\
        multiplexer out   & 12.7\hspace{0.3cm} & 22.9 & (+2.5) & 17.9 & (+2.0) & 15.7 & (+1.1)\\
        pipeline register & \textbf{13.6}\hspace{0.3cm} & \textbf{23.8} & (+0.0) & \textbf{19.1} & (+0.3) & \textbf{16.7} & (+0.1)\\ \bottomrule
    \end{tabular}
    \label{tab:timing}
\end{table}

\bibliographystyle{IEEEtran}
\bibliography{style, references}

\begin{thebibliography}{10}
\providecommand{\url}[1]{#1}
\csname url@samestyle\endcsname
\providecommand{\newblock}{\relax}
\providecommand{\bibinfo}[2]{#2}
\providecommand{\BIBentrySTDinterwordspacing}{\spaceskip=0pt\relax}
\providecommand{\BIBentryALTinterwordstretchfactor}{4}
\providecommand{\BIBentryALTinterwordspacing}{\spaceskip=\fontdimen2\font plus
\BIBentryALTinterwordstretchfactor\fontdimen3\font minus \fontdimen4\font\relax}
\providecommand{\BIBforeignlanguage}[2]{{%
\expandafter\ifx\csname l@#1\endcsname\relax
\typeout{** WARNING: IEEEtran.bst: No hyphenation pattern has been}%
\typeout{** loaded for the language `#1'. Using the pattern for}%
\typeout{** the default language instead.}%
\else
\language=\csname l@#1\endcsname
\fi
#2}}
\providecommand{\BIBdecl}{\relax}
\BIBdecl

\bibitem{kulu_satellite_2024}
\BIBentryALTinterwordspacing
E.~Kulu, ``\BIBforeignlanguage{en}{Satellite {Constellations} - 2024 {Survey}, {Trends} and {Economic} {Sustainability}},'' in \emph{\BIBforeignlanguage{en}{{IAF} {Businesses} and {Innovation} {Symposium}}}.\hskip 1em plus 0.5em minus 0.4em\relax Milan, Italy: International Astronautical Federation (IAF), 2024, pp. 54--83. [Online]. Available: \url{http://www.proceedings.com/078383-0004.html}
\BIBentrySTDinterwordspacing

\bibitem{di_mascio_open-source_2021}
\BIBentryALTinterwordspacing
S.~Di~Mascio, A.~Menicucci, E.~Gill, G.~Furano, and C.~Monteleone, ``\BIBforeignlanguage{en}{Open-source {IP} cores for space: {A} processor-level perspective on soft errors in the {RISC}-{V} era},'' \emph{\BIBforeignlanguage{en}{Computer Science Review}}, vol.~39, p. 100349, Feb. 2021. [Online]. Available: \url{https://linkinghub.elsevier.com/retrieve/pii/S1574013720304494}
\BIBentrySTDinterwordspacing

\bibitem{rogenmoser_trikarenos_2025}
\BIBentryALTinterwordspacing
M.~Rogenmoser, P.~Wiese, B.~E. Forlin, F.~K. Gürkaynak, P.~Rech, A.~Menicucci, M.~Ottavi, and L.~Benini, ``Trikarenos: {Design} and {Experimental} {Characterization} of a {Fault}-{Tolerant} 28-nm {RISC}-{V}-{Based} {SoC},'' \emph{IEEE Transactions on Nuclear Science}, vol.~72, pp. 2783--2792, Aug. 2025. [Online]. Available: \url{https://ieeexplore.ieee.org/document/10978878}
\BIBentrySTDinterwordspacing

\bibitem{bertozzi_error_2005}
\BIBentryALTinterwordspacing
D.~Bertozzi, L.~Benini, and G.~De~Micheli, ``\BIBforeignlanguage{en}{Error control schemes for on-chip communication links: the energy-reliability tradeoff},'' \emph{\BIBforeignlanguage{en}{IEEE Transactions on Computer-Aided Design of Integrated Circuits and Systems}}, vol.~24, pp. 818--831, Jun. 2005. [Online]. Available: \url{http://ieeexplore.ieee.org/document/1432874/}
\BIBentrySTDinterwordspacing

\bibitem{radetzki_methods_2013}
\BIBentryALTinterwordspacing
M.~Radetzki, C.~Feng, X.~Zhao, and A.~Jantsch, ``Methods for fault tolerance in networks-on-chip,'' \emph{ACM Computing Surveys}, vol.~46, pp. 8:1--8:38, Jul. 2013. [Online]. Available: \url{https://doi.org/10.1145/2522968.2522976}
\BIBentrySTDinterwordspacing

\bibitem{bhamidipati_hren_2022}
\BIBentryALTinterwordspacing
P.~Bhamidipati and A.~Karanth, ``{HREN}: {A} {Hybrid} {Reliable} and {Energy}-{Efficient} {Network}-on-{Chip} {Architecture},'' \emph{IEEE Transactions on Emerging Topics in Computing}, vol.~10, pp. 537--548, Apr. 2022. [Online]. Available: \url{https://ieeexplore.ieee.org/document/9703196}
\BIBentrySTDinterwordspacing

\bibitem{andorno_radiation-tolerant_2023}
\BIBentryALTinterwordspacing
M.~Andorno, A.~Caratelli, D.~Ceresa, J.~Dhaliwal, K.~Kloukinas, A.~Nookala, and R.~Pejasinovic, ``Radiation-{Tolerant} {SoC} and {Application}-{Specific} {Processors} for {On}-{Detector} {Programmability} and {Data} {Processing} in {Future} {High}-{Energy} {Physics} {Experiments},'' in \emph{2023 12th {International} {Conference} on {Modern} {Circuits} and {Systems} {Technologies} ({MOCAST})}, Jun. 2023, pp. 1--5. [Online]. Available: \url{https://ieeexplore.ieee.org/document/10176589/}
\BIBentrySTDinterwordspacing

\bibitem{arm_amba-apb_2021}
\BIBentryALTinterwordspacing
{ARM}, ``\BIBforeignlanguage{en}{{AMBA}-{APB} {Protocol} {Specification}},'' Apr. 2021, version D. [Online]. Available: \url{https://developer.arm.com/documentation/ihi0024/d/?lang=en}
\BIBentrySTDinterwordspacing

\bibitem{bink_obi_2023}
\BIBentryALTinterwordspacing
A.~Bink, ``{OBI},'' Mar. 2023, version 1.6.0. [Online]. Available: \url{https://github.com/openhwgroup/obi/blob/072d9173c1f2d79471d6f2a10eae59ee387d4c6f/OBI-v1.6.0.pdf}
\BIBentrySTDinterwordspacing

\bibitem{arm_amba_2023}
\BIBentryALTinterwordspacing
{ARM}, ``\BIBforeignlanguage{en}{{AMBA} {AXI} {Protocol} {Specification}},'' Sep. 2023, version K. [Online]. Available: \url{https://documentation-service.arm.com/static/651c285c15583d1bff972f94}
\BIBentrySTDinterwordspacing

\bibitem{kulis_single_2017}
\BIBentryALTinterwordspacing
S.~Kulis, ``Single {Event} {Effects} mitigation with {TMRG} tool,'' \emph{Journal of Instrumentation}, vol.~12, pp. C01\,082--C01\,082, Jan. 2017. [Online]. Available: \url{https://iopscience.iop.org/article/10.1088/1748-0221/12/01/C01082}
\BIBentrySTDinterwordspacing

\end{thebibliography}

\end{document}